\documentclass[showpacs,prl,twocolumn]{revtex4}
\usepackage{graphicx}
\usepackage{dcolumn}
\usepackage{bm}
\begin{document} 

\title{Local driving and global interactions in the progression of seizure dynamics}

\author{Benjamin H. Singer, Miron Derchansky, Peter L. Carlen, and  Micha{\l}\ \.Zochowski}
\affiliation{Neuroscience Program, Department of Physics and Biophysics Research Division\\
University of Michigan\\
Ann Arbor, MI 48109}
\affiliation{Division of Cellular and Molecular Biology\\
Toronto Western Research Institute, Ontario\\
Toronto, Canada}

\begin{abstract}
The dynamics underlying epileptic seizures are well understood. We present a novel analysis of seizure-like events (SLEs) in an \textit{ex vivo} whole hippocampus, as well as a modeling study that sheds light on the underlying network dynamics. We show that every SLE can be divided into two phases. During the first, SLE dynamics are driven by the intra-network interaction of a network exhibiting high internal synchrony. The second phase is characterized by lead switching, with the leading region exhibiting low internal synchrony.  We show that the second phase dynamics are driven by inter-network feedback among multiple regions of the hippocampus.
\end{abstract} 
\date{\today}
\pacs{87.18.Hf, 05.45.Xt, 05.65.Tp}
\maketitle 

Epilepsy is a disease that occurs in 2\% of the population and is though to be caused by interplay of variety of molecular, cellular and network mechanisms\cite{avoli}. However, the common underlying feature of all epilepsies is reoccurring seizures due to spontaneous indiscriminate synchronization and bursting of cell populations \cite{dudek99}. Thus, it is crucial to understand the fundamental properties of seizure dynamics in order to reduce their occurrence. 

The hippocampus is thought to play a central role in the genesis of seizures in the most common form of epilepsy \cite{wasterlain96}.   While many studies have elucidated the dynamics of seizure onset \cite{schiff02,lehnertz03,netoff04,lds03}, less in known about internal seizure dynamics.  Here, we use a novel technique to analyze seizure-like events (SLEs) in an \textit{ex vivo} preparation of the whole hippocampus.  We have found that the observed seizure-like activity can be divided into two phases. During the first phase SLE dynamics are driven by the region of the hippocampus which exhibits the highest internal synchrony.  Using a computational model, we propose that in the first phase the intra-network dynamics of a single region drive the SLE throughout the hippocampus.  The second phase is marked by the onset of lead switching among regions of the hippocampus and, contrary to the first phase, the leading region has the lowest internal synchrony.  We attribute this inversion of the relationship between leading region and local synchrony to SLE dynamics reflecting inter-network feedback among hippocampal regions. 

\begin{figure}
	
		\includegraphics[width=3.25in]{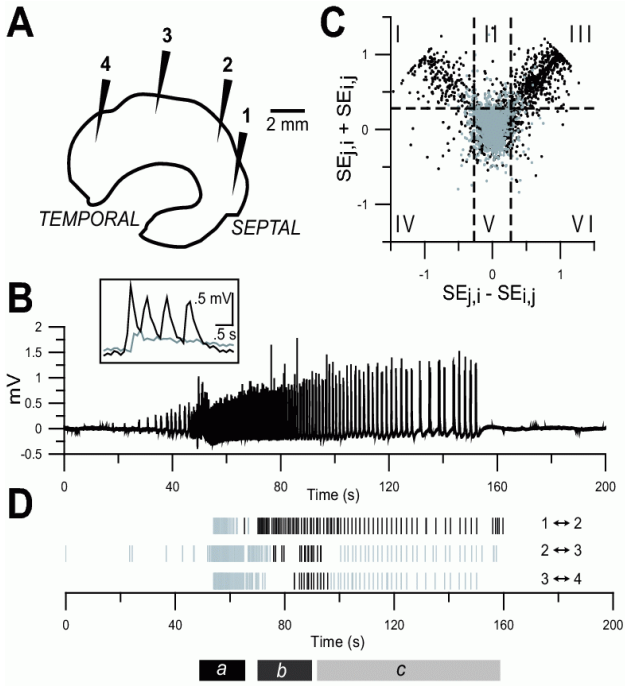} 
		\caption{\footnotesize{
\textit{A:} Schematic of the hippocampus showing the temporal and septal poles and the approximate placement of electrodes. 
\textit{B:} Representative local field potential recording of a single SLE (electrode 2).  \textit{Inset:} a single burst from mid-SLE, showing electrode 2 (black) leading electrode 1 (gray).
\textit{C:} Two dimensional plot of $SE_{j,i} +SE_{i,j}$ vs. $SE_{j,i} -SE_{i,j}$ for data shown in \textit{B} (black).  Values of $SE_{i,j}$ are paired with their nearest temporal neighbors in  $SE_{j,i}$.  Thresholds for SE sum and difference are set based on the distributions for a randomly shuffled surrogate dataset (gray).  Thresholds are set at the 95$^{th}$ percentile of $SE_{j,i} +SE_{i,j}$ (horizontal line), and the 2.5$^{th}$ and 97.5$^{th}$percentiles of  $SE_{j,i} -SE_{i,j}$ (vertical lines).  The intersection of these thresholds form six sectors.  In sector I, activity in $i$ significantly leads activity in $j$, while in sector III,  activity in $j$ significantly leads activity in $i$. Events in other sectors are not signficant.
\textit{D:} Raster plots showing times of events pairs among adjacent electrodes $\{(1,2), (2,3), (3,4)\}$with significant SE sum and difference values.  Color denotes septal electrode leading temporal (gray) and temporal electrode leading septal (black).  The overall pattern of leading changes from septal $\rightarrow$ temporal (\textit{a}) to temporal $\rightarrow$ septal (\textit{b}) to center $\rightarrow$ poles (\textit{c}).
}}
		\label{fig:recordings}
\end{figure}

Seizure activity in animal models of epilepsy has long been studied in transverse hippocampal slices \cite{avoli}.  However, slicing the hippocampus creates artificial two dimensional neuronal circuits which do not capture dynamics resulting from the three dimensional organization of the hippocampus.  Here, spontaneous seizure-like activity is studied by field potential recording from whole hippocampi dissociated from the brains of C57/BL mice, as described in \cite{carlen}.  The excitability of the neural tissue is increased by superfusing with low Mg$^{2+}$ artificial cerebrospinal fluid, thus inducing spontaneous, recurrent SLEs.  Four electrodes are placed along the temporal-septal axis of the hippocampus (Fig.\ \ref{fig:recordings}A).    
Recordings from the four electrodes thus represent neural activity in widely separated regions along the longitudinal hippocampal axis, and reflect both network dynamics within the local networks of the four regions, and inter-network interactions among them.  Bursts recorded in these four regions are highly coincident with relatively small lead  and width variations.  Our goal is to characterize how the interaction among distinct regions along the temporal-septal axis relate to the evolution and termination of individual SLEs.

\begin{figure}

	\includegraphics[scale=1.0]{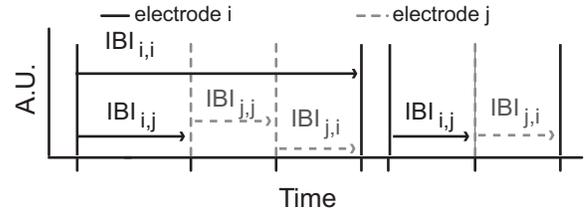}
 
		\caption{\footnotesize{
The interval $IBI_{i,j}$ of the $i$-th electrode with respect to the $j$-th electrode is calculated as the time difference between an event in $j$ and the immediately preceding event in $i$.  $CE_{i,j}$ is updated after every calculation of the running distribution of $IBI_{i,j}$, and assigned to the time of the originating event in $i$. Analogous calculations carried out for $CE_{j,i}$, $S_{i,i}$, and $S_{j,j}$.  
}}
		\label{fig:CE}\end{figure}

Recordings from the \textit{ex vivo} hippocampus reveal abrupt transitions from interictal periods to SLEs, which are composed of slow bursts and superimposed fast spikes (Fig.\ \ref{fig:recordings}B).  
To discern the dynamic relationship among the four recorded regions, we examine the small temporal variations in the onset of slow bursts (Fig.\ \ref{fig:recordings}B, inset), as recorded in the four electrodes.  We use a modified version of a measure we have ealier developed \cite{mrzrd1-03} to quantify differences in relative inter-burst intervals (IBIs) between electrode pairs (Fig.\ \ref{fig:CE}), with the underlying idea that activity in a driven region will closely and systematically follow the activity of the driving region with minimally variable IBIs .  The IBIs are calculated separately for every electrode pair $(i,j)$ in the the network, and distributions of  IBI values are constructed.  The running distributions of relative IBIs ($IBI_{i,j}$)  are updated and renormalized over time with each the onset of each burst.  After each update, the Shannon entropy, $S=\sum_{k}p_k \ln{p_k}$,  of the renormalized IBI distribution is calculated. This quantity depends on the relative timing of bursts in both electrodes in the pair, and is referred to as a conditional entropy of electrode $i$ vs. electrode $j$ ($CE_{i,j}$).

Since the relative IBIs are measured unidirectionally (Fig.\ \ref{fig:CE}), the pairwise comparison of $CE_{i,j}$ and $CE_{j,i}$  allows the asymmetric measurement of temporal interdependencies between the activity at two electrodes.  To account for the possibility that the temporal interdependence between two signals could be an artifact of autonomous signal properties on one (or both) electrodes we calculate the quantity:
\begin{equation}
SE_{i,j}=\frac{min(S_{i,i}, S_{j,j})-CE_{i,j}}{min(S_{i,i},S_{j,j})},
\label{SE}
\end{equation}
where $S_{i,i}$ is the entropy of (continuously updated) IBI distributions observed on a single electrode ($IBI_{i,i}$) . The values of SEs around and below zero indicate that there is no significant interdependence between the signals, whereas if SE tends to one this indicates strong temporal interdependence.  This measure detects and categorizes three basic regimes: the two electrodes are independent (both $SE_{i,j}$ and $SE_{j,i}$ are low); the signal at electrode $i$ leads electrode $j$ ($SE_{i,j}$ is high while $SE_{j,i}$ is low);  the signal is nearly periodic or is completely synchronized (both $SE_{i,j}$ and $SE_{j,i}$ are high). 

SE values are computed for every burst of an SLE.  In order to determine which SE values are due to underlying neural activity and which are due to chance, we compute surrogate datasets by randomly reassigning burst times from an individual recording to new electrode labels (shuffled dataset).  Thresholds for significance are then based on the distribution of $(SE_{j,i}+SE_{i,j})$ and $(SE_{j,i}-SE_{i,j})$ over 10 shuffled datasets (Fig.\ \ref{fig:recordings}C).  
Events that fall outside the significance threshold for both SE sum and difference for each of the electrode pairs $(i,j) = \{(1,2), (2,3), (3,4)\}$ are depicted in a raster plot of individual events and lead pattern (Fig.\ \ref{fig:recordings}D).  Within the SLE depicted, the focus of activity is initially stable and located in the septal (electrode 1) region of the hippocampus (Fig.\ \ref{fig:recordings}D, region \textit{a}).  The lead pattern rapidly switches so that temporal (electrode 4) activity leads septal activity (Fig.\ \ref{fig:recordings}D), region \textit{b}).  The lead pattern then switches so that a region in the mid-hippocampus (electrode 2) leads both septal and temporal poles, and this pattern remains stable until the SLE ends (Fig.\ \ref{fig:recordings}D, region \textit{c}).  We find that this pattern of septal, temporal, and central lead transitions is remarkably stable over different SLEs recorded from different hippocampi .   This pattern is observed in 19/23 (82 \%) of SLEs analyzed, recorded from 6 hippocampi.  

\begin{figure}
	
			\includegraphics[scale=1.0]{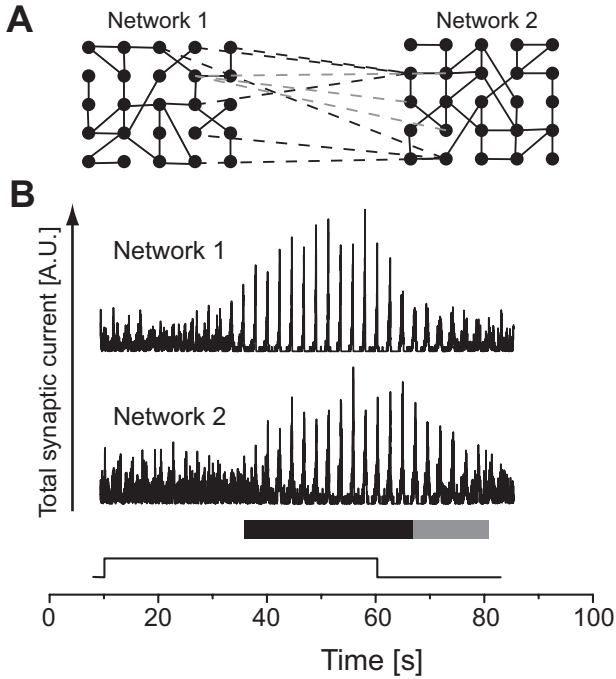} 
		\caption{\footnotesize{
\textit{A:}  Schematic connectivity of two connected SWN networks.
\textit{B:} Total synaptic activity from each network.  In the first phase of the model seizure, activity in N1 leads activity in N2 (black bar).  In the second phase, switching occurs and activity in N2 leads activity in N1 (gray bar).  Step function denotes the period ($t \in [10,60]$) when $\Gamma_{cut} = 0$.  
}}
		\label{fig:model}
\end{figure}

In order to further elucidate this observed switching behavior, we build a simple computational model capturing basic properties of the experimental system.  We assume that each region recorded by a single electrode constitutes a local network which is interconnected with other networks along the temporal-septal axis of the hippocampus.  To model the simplest case of two such interconnected networks, we create two networks of integrate-and-fire neurons with a Small World Network (SWN) architecture ($N=15x15$ grid each, Fig.\ \ref{fig:model}A).  The SWN architecture constitutes an intermediate type of connectivity between local and global connectivity \cite{strogatz}, has been reported in neural structures, and has been linked to seizure generation \cite{netoff04}.  The neurons in both networks are positioned on a 2-dim lattice with a lattice constant $a=1$ and periodic boundary conditions.  Initially, in each network all the neurons within a radius $k=2$ are connected.  A fraction ($p=0.3$) of those initial connections are then randomly rewired.  This results in a networks having $0.05$ connectivity ratio.  Additionally, a fraction, ($f=0.3$), of randomly chosen neurons in one network receives the synaptic current from randomly chosen group of $m = 10$ neurons from the other network.  The dynamics of each neuron are given by:
\begin{equation}
\frac{dV_i}{dt}=-\alpha_i V(t) +A\sum_{j\in C} J_{ij}(t) + B\sum_{k\in I} J_{ik}(t) + \xi(t)
\label{IFneuron}
\end{equation}
where $A=4$ determines the intra-network signal amplitude, $B=0.4$ is the inter-network signal amplitude, $\alpha_i\in(1.0, 1.5)$ is the membrane leakage coefficient (different for every neuron in the network),  and $\xi\in(0.0,1.4)$ is a random variable simulating white noise. $C$ denotes the set of all neurons connected to $i$-th neuron via \emph{intra}-network connections, while $I$ denotes the neurons connected via \emph{inter}-network connections. 
$J_{ij}$ is the term describing synaptic current arriving from the $j$-th neuron and is given by:
\begin{equation}
J_{ij}=(\exp{(-\frac{t_s}{\tau_s})}-\exp{(-\frac{t_s}{\tau_f})})
\label{synapse}
\end{equation}
where $t_s$ is the time from the last spike generated at $j$-th neuron;  $\tau_{s}=0.3$ms is a slow time constant, whereas $tau_f=0.03$ms is a fast time constant. The interplay of those two constants defines the time course of the spike decay.
  
When the threshold $\Gamma_{spike}=1$ is reached, a spike is generated and the membrane potential is reset to $0$. During a built-in post-spike refractory period, $T=10$ms, the membrane does not potentiate in response to incoming stimuli.
Every neuron in both networks has an additional inhibitory mechanism that resets the incoming synaptic current to zero if it is below a threshold level ($\Gamma_{cut}=0.3$).  Inclusion of this threshold imposes a requirement for coincident input in spike generation, much as in dendritic processing.
The parameters if the model ($p$, $A$, $B$, $\xi$) are set so that both networks are just below a spontaneous bursting regime. At a set point in time ($t=10$s, Fig.\ \ref{fig:model}B, bottom) $\Gamma_{cut}$ is set to zero in network 1, shifting network 1 into spontaneous bursting.  The change in $\Gamma_{cut}$ is a phenomenological model of a transiently lowered firing threshold, which could be due to multiple neurobiological mechanisms.  For example, on the single-cell level it could be driven by changes in Ca$^{2+}$ dynamics and loss of hyperpolarizing activity from Ca$^{2+}$-dependent K$^+$ channels \cite{gorter}.

Bursting in network 1 (N1) is generated when sufficient numbers of neurons spike simultaneously, generating a cascade effect in the network.  Thus bursting in network 1 is initially generated through intra-network dynamics following a delay after lowering $\Gamma_{cut}$.  The activity in N1, in turn, provides input to network 2 (N2), resulting in seizure-like activity in both networks.  Bursting in N2 is therefore a result of synchronous, inter-network signalling, the pattern of which is dictated by the internal dynamics of N1.  In this phase of the model seizure, when $\Gamma_{cut}$ is low, the bursts of N1 lead those of N2 (Fig.\ \ref{fig:model}B, black bar).  N1 bursts are also generally higher and narrower than N2 bursts, indicating greater internal synchrony of N1.

At $t=60$s $\Gamma_{cut}$ in N1 is reset to its original value. As a result, the dynamics of the 2 network system changes dramatically. Bursting dynamics are no longer due to an increased intrinsic firing rate of neurons in N1, but are sustained by synaptic input alone.  Thus, the bursting activity of both networks is not predominantly mediated by the internal dynamics of N1, but by inter-network feedback. In this phase, switching in temporal leading and internal synchrony are observed among the two networks, due to an inverse driving phenomenon (Fig.\ \ref{fig:recordings}B, gray bar).   
The instantaneous rate of bursting in each network is inversely linked to the size and coherence of the bursts generated by the other, consistent with bursting sustained by inter-network interactions  For example, if a low-amplitude, diffuse burst in N1,  will result in less activation of N2.  This, in turn, leads to decreased instantaneous frequency and gradual desynchronization of N2.  Due to the slowing of N2 bursts, N1 activity will temporally lead bursting in N2.  Thus, the less synchronous and smaller bursts of one network will lead the more synchronous and larger bursts of the other until bursting ceases.

\begin{figure}
			\includegraphics[scale=1.1]{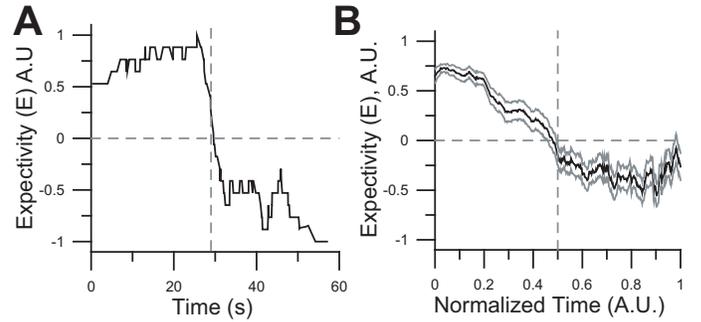} 
		\caption{\footnotesize{
\textit{A:} Average of expectivity among 10 simulated seizures, as defined in the text.  Left of dashed vertical line: $\Gamma_{cut} = 0.0$.  Right of dashed vertical line: $\Gamma_{cut} = 0.3$.   
\textit{B:} Average (black) $\pm$ SEM (gray) of expectivity ($E$) among 45 electrode pairs, as defined in the text.  Time from SLE onset to switching is normalized to $[0,0.5]$, and time from switching to SLE end is normalized to $[0.5,1.0]$.  \textit{Vertical line:} onset of switching.      
}}
		\label{fig:expectivity}
\end{figure}

To quantify this inverse driving effect, we compare the ratio $\gamma=\frac{height}{width}$ of bursts in each network to the identity of the leading network for 10 simulated seizures of the type shown in (Fig.\ \ref{fig:model}B).  To measure relationship between $\gamma$ and temporal leading, we define the expectivity of a burst as $+1$ if the leading burst has a greater value of $\gamma$ than the following burst, and $-1$ if the leading burst is smaller in $\gamma$ than the following burst.  In order to capture the temporal pattern in expectivity and control for variations in burst freqency, we average expectivity over a sliding window.  In the simulated seizures, expectivity is positive while $\Gamma_{cut} = 0$ in N1, and undergoes a rapid transition to negative values when $\Gamma_{cut}$ is restored (Fig.\ \ref{fig:expectivity}A).  This finding is consistent with the transition from driving by intra-network dynamics to bursting sustained by inter-network feedback.

To determine if similar dynamics may be observed in recordings from the whole hippocampus, we define expectivity among two electrodes as:
\begin{equation}
E_{i,j} = \left\{ 
\begin{array}{ll}
+1, & (W_i - W_j)(SE_{j,i}-SE_{i,j}) \geq 0 \\
-1, & (W_i - W_j)(SE_{j,i}-SE_{i,j}) < 0
\end{array} \right\}
\label{eq:expect}
\end{equation}
where $W$ is the burst width.
Expectivity ($E_{i,j}$) is calculated on a burst-by-burst basis and averaged over sliding windows for each adjacent electrode pair.  We use burst width as a measure of intra-network synchrony, since it is less confounded by current source location than burst amplitude.  In order to account for the variable length of SLEs across recordings, we linearly map the time from SLE onset to switching onset to the interval $[0,0.5]$, and the time from switching onset to the end of the SLE to the interval $[0.5, 1.0]$.  These normalized time series are then averaged over 45 comparisons of activity in adjacent electrodes to calculate average expectivity ($E$).  We observe that before the onset of switching ($t<0.5$), $E > 0$, and the electrode displaying narrower bursts leads the pair (Fig.\ \ref{fig:expectivity}B, left).  After the onset of switching ($t>0.5$), $E < 0$, with wider bursts leading (Fig.\ \ref{fig:expectivity}B, right), as observed in our model.  
Thus, these findings are consistent with a transition from the intra-network dynamics of a single region driving activity throughout the hippocampus to bursting sustained by inter-network feedback, as observed in our model.   

To our best knowledge this is the first such recording, characterization, and proposed mechanism for the dynamics of SLE in the whole hippocampus.  Using a novel measure, we observe and characterize lead switching among regions of the hippocampus over the course of an SLE. We contruct a simple model to provide insight into the dynamics underlying these observations.  We show that each SLE is divided into two phases. The first phase is driven by local intra-network dynamics of the region with greatest local synchrony, while the second phase is characterized inter-network feedback and an inverse leading relationship.

B.S. and M.Z. thank Professor Eshel Ben-Jacob, Professor Geoff Murphy, and Professor Jack Parent.  This work was supported by a UM Research Incentives Grant (M.Z.) and CIHR research grant MT14447 (P.L.C).  B.S. is supported by the UMMS Medical Scientist Training Program (NIH T32-GM007863).


\end{document}